\documentclass{elsart}
\usepackage{graphicx}
\usepackage{amssymb}
\journal{Solid State Communications}
\begin{document}
\begin{frontmatter}
\title{Field induced first order antiferromagnetic to ferromagnetic transition in Al-doped CeFe$_2$:  a calorimetric investigation}
\author{L.S. Sharath Chandra, Kaustav Mukherjee, V. Ganesan,}
\author{A. Banerjee and P. Chaddah}
\address{UGC-DAE Consortium for Scientific Research (CSR),
University Campus, Khandwa Road,
Indore 452017, India.}
and
\author{M.K. Chattopadhyay and S.B. Roy}
\address{Magnetic and Superconducting Materials Section, Centre for Advanced Technology, Indore 452013, India.}
\date{\today}
\begin{abstract}Field variation of heat capacity at fixed temperatures is investigated to identify the origin of the field induced first order phase transition in polycrystalline  Ce(Fe$_{0.094}$Al$_{0.04}$)$_2$ sample.  The heat capacity at 4.5K and 3.5K shows hysteresis in different field cycles and the virgin curve stays outside the envelope curve. This is analogous to the magnetization and magneto-resistance behavior observed in this system, where the amount of hysteresis, and the magnitude of zero-field irreversibility are attributed to the degree of supercooling/superheating, and the extent of kinetic arrest of the reverse transition from ferro- to antifrromagnetic state in field reducing cycle respectively.   However, contrary to the magnetization and magneto-resistance, in heat capacity both these features have decreased with reducing temperature signifying the importance of structural contribution.  
\end{abstract} 
\end{frontmatter}
\maketitle
\section{Introduction}
The C15 laves phase ferromagnet CeFe$_2$ has an intrinsic magnetic instability in the form of antiferromagnetic fluctuations at low temperatures\cite{1}.  Such antiferromagnetic fluctuations get stabilized with small dopings of selective elements like Co, Ru, Al, Ir and Os at the Fe site, and a long range antiferromagntic state develops at low temperatures\cite{2}. This ferromagnetic (FM) to antiferromagnetic (AFM) transition in doped-CeFe$_2$ alloys has drawn quite a bit of attention during last two decades, and the first order nature of this transition has been established through various measurements\cite{3,4,5,6}. Quenched random disorder plays a crucial role on this first order transition in doped CeFe$_2$ alloys. The characteristic features have been identified, and the generality of the phenomena pointed out, by comparing with the similar  features observed  in various other family of compounds including CMR manganites and magnetocalric material Gd$_5$Ge$_4$\cite{7}. There are additional interesting features mainly observed in Al-doped CeFe$_2$ alloys which indicate that in certain field -temperature  regime the first order transition process is kinetically arrested\cite{6}. This arrested state has the characteristic feature of a glass, namely it is non-ergodic in nature. In field variation measurements, this kinetic arrest gives rise to an interesting irreversibility where the virgin AFM state (obtained after zero field cooling from high temperature) cannot be regained completely after a field reduction from the field-induced FM state. Such zero-field irreversibility gives rise to anomalous isothermal magnetization (M) and resistivity ( R ) versus  field (H) curves, where the virgin M(or R) versus H curve lies distinctly outside the envelope curve\cite{6}. In this communication we investigate this  field induced AFM to FM transition process, and the kinetic arrest of the field reversed FM to AFM transition, with calorimetric technique.  The results of the present specific heat measurement clearly indicate that the structural change associated with the first order AFM-FM transition plays a crucial role in the low temperature kinetic arrest, as well. 

\section{Experimental}
The small ( 5 mg) 4\%Al-doped CeFe$_2$ sample used in this study is taken from 4 gm ingot obtained by argon arc melting from 99.99\% purity starting materials. The sample has been subsequently annealed at 800 \raisebox{1ex}{\scriptsize o}C for 7 days. The main mass of the sample has been used for neutron diffraction measurements\cite{8}. This measurement has indicated the presence of a small amount (less than 5\%) of Ce$_2$Fe$_{17}$ impurity phase. This Ce$_2$Fe$_{17}$ impurity phase adds a ferromagnetic component in the field dependent magnetization of the low temperature AFM state. However, it does not influence the field induced AFM-FM transition which is the main subject of the present work.  The specific heat measurements in the magnetic field up to 14 Tesla have been performed using a commercial setup (14 Tesla PPMS, Quantum Design). The sample platform used for measurements above 2K is temperature as well as field calibrated and the addenda contributions are appropriately subtracted from the measured total specific heat.  The typical error in the sample specific heat is estimated to be $<$0.5\%.  A He3 insert for the PPMS is used for the measurement at 800 mK. The addenda contributions are not subtracted from the total specific heat in this case.  This raw data is used for the limited purpose of estimating the extent of zero-field irreversibility at a much lower temperature.  For all the measurements proper care is taken not to overshoot/undershoot either in temperature or in field when the respective parameter is stabilized.

\section{Results and Discussion}
\begin{figure}[p]
	\centering
		\includegraphics{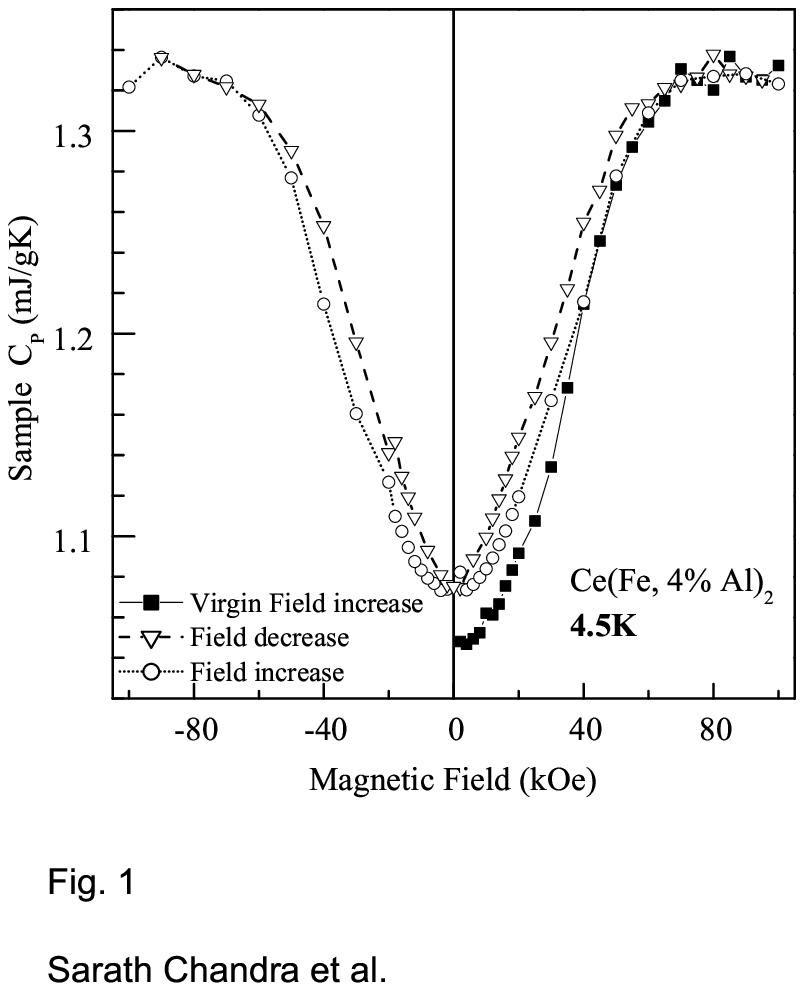}
	\caption{Specific heat of Ce(Fe, 4\%Al)$_{2}$ as  a function of magnetic field at 4.5 K.  The specific heat is measured after zero-field cooling from 300K to 4.5K.  The initial or the virgin field increasing cycle (filled square) and the field decreasing (open triangle) as well as field increasing (open circle) cycles for both +ve and -ve field excursions are shown.}
	\label{fig1}
\end{figure}

\begin{figure}[p]
	\centering
		\includegraphics{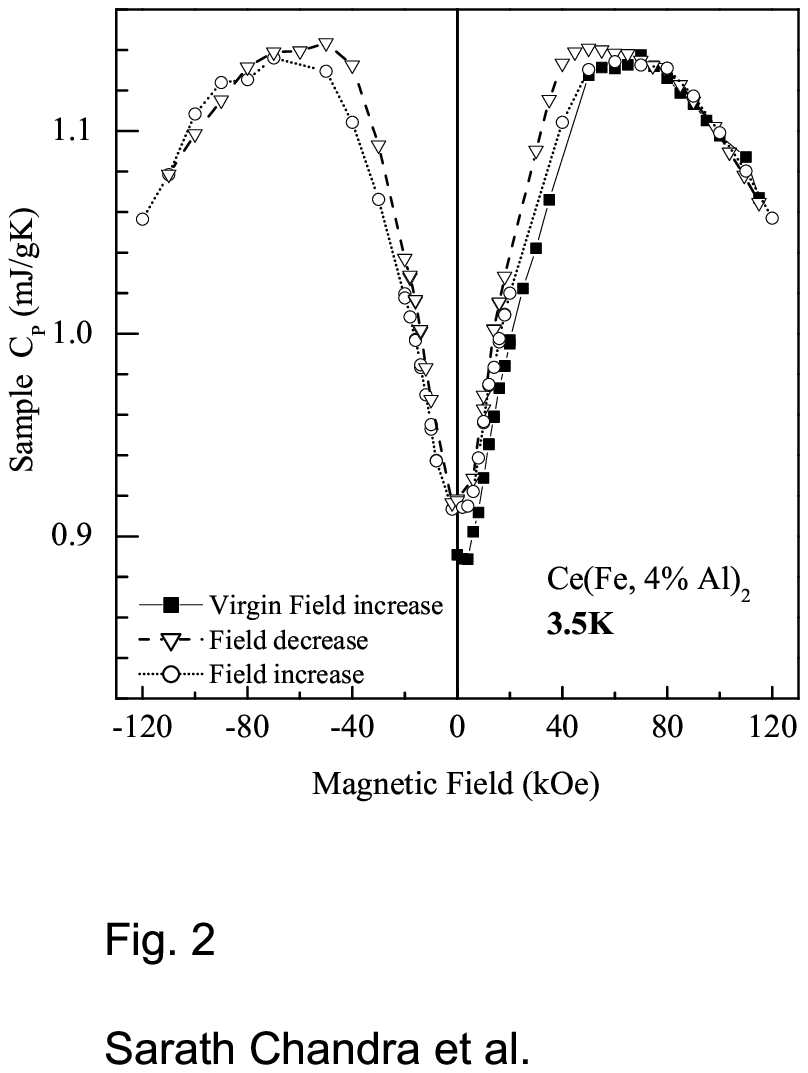}
	\caption{Specific heat of Ce(Fe, 4\%Al)$_{2}$ as  a function of magnetic field at 3.5 K.  The specific heat is measured after zero-field cooling from 300K to 3.5K.  The initial or the virgin field increasing cycle (filled square) and the field decreasing (open triangle) as well as field increasing (open circle) cycles for both +ve and -ve field excursions are shown.}
	\label{fig2}
\end{figure}

Figure ~\ref{fig1} shows the specific heat (C$_P$) versus field (H) plot of  the sample at 4.5K. The sample is zero field cooled (ZFC) to 4.5 K from 300K before the magnetic field is switched on. C$_P$  increases with H, and there is a marked change in slope in the C$_P$ vs H curve at 35 kOe. This field coincides with the onset of AFM-FM transition obtained from the isothermal field dependent magnetization measurements at 4.5 K\cite{6}. The C$_P$  versus H curve tends to saturate above 70 kOe. This field  roughly matches  with the field value where the formation of  FM state was complete in the  isothermal field variation of magnetization. On returning from 100 kOe the C$_P$ versus H curve separates from the virgin curve roughly below 40 kOe and the zero field C$_P$  remains distinctly above the virgin zero field value. We have done a negative filed excursion to 100 kOe and the field is subsequently increased to 100 kOe .   This leads to a roughly symmetric envelope C$_P$ vs H curve, and the virgin  C$_P$ vs H curve lies distinctly outside this envelope curve. Similar behaviour is observed at 3.5K which also shows a difference between virgin C$_P$(H=0) and remanent C$_P$(H=0) values (see Figure ~\ref{fig2}).  It has been argued earlier that such anomalous virgin curve arises due to the kinetic arrest of the FM to AFM transition process in the descending field cycle\cite{6}. We shall argue here that the kinetic arrest does not actually cause the freezing of  spin degrees of freedom. The kinetic arrest involves freezing of structural configurations and the corresponding amount of the respective long-range ordered phase, in this case high-temperature/high-field ferromagnetic phase.   

\begin{figure}[p]
	\centering
		\includegraphics{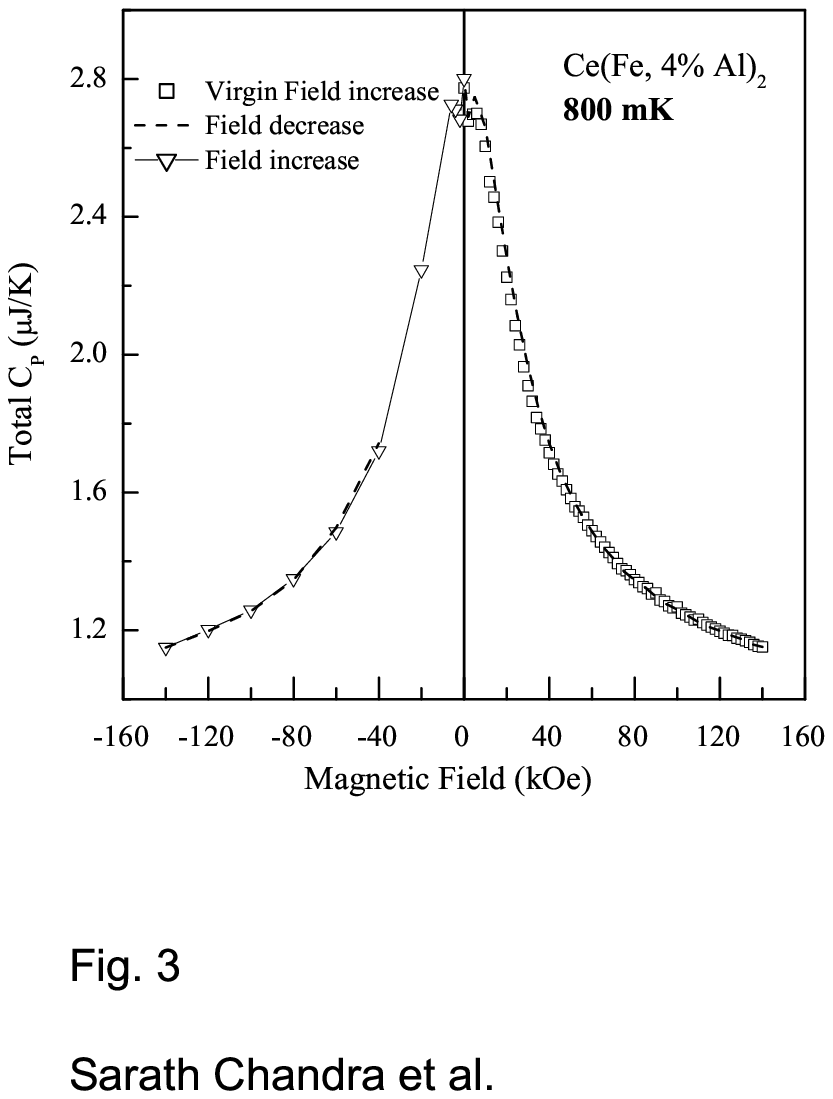}
	\caption{Total specific heat of Ce(Fe, 4\%Al)$_{2}$ alongwith the addenda as function of magnetic field at 800mK. The specific heat is measured after zero-field cooling from 300K to 800 mK.  The initial or the virgin field increasing cycle (open square) and the field decreasing (dash line) as well as field increasing (open trangle) cycles for both +ve and -ve field excursions are shown.  The pronounced  hysteresis  and signature of kinetic arrest observed at higher temperatures are absent here.}
	\label{fig3}
\end{figure}

We note in the C$_P$ vs H curve at 3.5 K (see Figure ~\ref{fig2}), there is a distinct down turn in the high field region. Also the field induced AFM to FM transition is not clearly discernable. It is well known that there are two distinct sources of contributions to specific heat in metallic systems -from conduction electrons and phonons. Due to the intermediate valence nature of 4f electrons in CeFe$_2$ the electronic contribution to specific heat is quite high \cite{9}. In low temperature region below 4K it is expected to be comparable to phononic contribution, and start dominating at certain stage.  We also know that in various intermediate valence and heavy fermion compounds the electronic coefficient of specific heat gets suppressed in applied magnetic field in the temperature region roughly below 4K\cite{10,11}. We believe that all these effects combine together to give rise to the high field downturn in the C$_P$ vs H curve at 3.5K. This conjecture is further strengthened by the observed field dependence of C$_P$  at 800mK (see Figure ~\ref{fig3}).  Neither the metamagnetic transition nor any signature of kinetic arrest is visible in the C$_P$ vs H curve at 800mK. We argue that below 1K  electronic contribution is dominating the specific heat, and there is negligible contribution from phonons. We recall here that during the temperature induced FM to AFM transition there is a rhombohedral distortion of the higher temperature cubic phase\cite{3}. This high temperature cubic phase is restored during the field induced AFM to FM transition. It appears that the signature of  the AFM to FM transition in the specific heat measurements mainly comes from the change in phonon structure accompanying this rhombohedral to cubic structure restoration. While reducing the field from the high field cubic-FM state, the cubic to rhombohedral distortion is not quite complete due to the kinetic arrest of the transition process . Hence the phononic contribution to the zero field specific heat comes from a structural configuration consisting of part transformed rhombohedral-AFM state and untransformed cubic-FM state. This configuration is quite different from a total rhombohedrally distorted AFM state obtained in the ZFC path. This leads to the distinct difference between the ZFC and remanent specific heat. At lower temperatures the amount of untransformed  cubic-FM state will be higher during the kinetic arrest of the FM to AFM transition and it gives rise to larger difference between ZFC and remannent magnetization. This signature, however, is absent in the specific heat measurement because it is not sensing the FM state directly but through the characteristic phononic contribution coming from the cubic structure of the frozen FM state. Since the phononic contribution itself is dying down with the reduction of T, the signature of arrest also diminishes in calorimetric study at low temperatures.   This is reinforced by the hysteresis at 3.5K being less than at 4.5K, and the 800mK data showing no clear indication of either the transition or the arrest.

\section{Conclusion}
In conclusion we have investigated the field induced first order AFM to FM transition in 4\%Al-doped CeFe$_{2}$ alloy with calorimetric measurements. Clear signature of  kinetic arrest  of the reverse transition from FM to AFM state is observed in the field reducing cycle.  As in the data shown here in figures 1 and 2, the virgin zero-field value is not restored on the field-reducing cycle.  This zero-field irreversibility is belived to result from kinetic arrest.  The same characteristic signatures  have been observed in magnetic and magneto-transport properties during FM to AFM transition in various other families of magnetic materials \cite{12,13,14,15}, and a connection with the possible kinetic arrest has been identified in a few cases \cite{16,17}.
  In both magnetic and magneto-transport measurements, the extent of this zero-field irreversibility rises as temperature falls.  The reverse trend is observed in our calorimetric measurements.  This underscores that the irreversibility is structural in origin, because the relative structural or phononic contribution to specific heat fall rapidly with the fall of temperature.  We believe that all these results are understandable within the same generalized phenomenological framework of a first order phase transition and its kinetic arrest in the ceratin region of the field-temperature phase space.  It is also interesting to note here that a structural change is involved in all these magnetic materials . It seems that the strain disorder coupling is the key parameter which controls the transition process.  It will be interesting now to make a calorimetric investigation of  the FM to AFM transition in all these materials.
  
\section{Acknowledgements} We acknowledge Prof. A. Gupta for his support to establish the Low Temperature \& High Magnetic Field (LTHM) facilities at CSR, Indore Center.  Mr. Kranti Kumar and Mr. M. Gangrade are acknowledged for their help in installation and operation of the experimental facilities.   Mr. P. Saravanan and other members of the cryogenic section are acknowledged for their efforts in smooth operation of the systems.  Department of Science and Technology, Government of India, is acknowledged for its support in establishing the LTHM facilities at CSR, Indore Centre.

\newpage

\end{document}